\documentclass[12pt, a4paper, notitlepage]{article}

\usepackage{amssymb, amsthm}
\usepackage{amsmath}
\usepackage[utf8]{inputenc} 
\usepackage[T1]{fontenc}    
\usepackage{hyperref}       
\usepackage{url}            
\usepackage{booktabs}       
\usepackage{amsfonts}       
\usepackage{nicefrac}       
\usepackage{lipsum}		
\usepackage{graphicx}
\usepackage{titlesec}
\usepackage{cancel}
\usepackage{titling}
\usepackage [autostyle, english = american]{csquotes}
\usepackage{changepage}
\usepackage{esvect}
\usepackage{enumitem}

\renewenvironment{abstract}
 {\small
  \begin{center}
  \bfseries \abstractname\vspace{-.5em}\vspace{0pt}
  \end{center}
  \list{}{
    \setlength{\leftmargin}{.3cm}%
    \setlength{\rightmargin}{\leftmargin}%
  }%
  \item\relax}
 {\endlist}

\MakeOuterQuote{"}

\title{Path Integral Formulation and Holonomy Groups in Newton-Cartan Schwarzschild Geometry}

\author{Kristo Nugraha Lian \\
	Department of Mathematics\\
	National University of Singapore\\
	Singapore, 119077 \\
	\texttt{kristolian@yahoo.com} \\
}

\date{\today}

\newenvironment{widerequation}{%
    \begin{adjustwidth}{-3cm}{-2cm}\begin{equation}}
    {\end{equation}\end{adjustwidth}}

\begin{document} 	
\begin{titlingpage}
    \maketitle
    \begin{abstract}
We use vielbein bundle's horizontal lift path integral formulation and gauge theory's holonomy map to compactly describe parallel transport and geodesic equations on a manifold.  This is first applied to the geometry of general relativistic Schwarzschild as a review.  The Newton-Cartan take of the Schwarzschild metric derived by previous literature is then adopted, and the analysis is repeated for both non-torsional and torsional geometry.  Transport curves considered include a circular timeless loop, circular geodesic loop, radial geodesic loop, and a stationary loop.  The findings on the three geometries are then contrasted, key differences summarized and the differing mechanisms discussed.  In particular, we find a "performance tradeoff" between the torsional and non-torsional Newton-Cartan theory: the former yields more accurate equations of motion whereas the latter simulates relativistic dilation/contraction effects to an exact degree.
    \end{abstract}
\end{titlingpage}

\section{Introduction}\label{sec1}

\paragraph{}
Albeit being a classical approximation of general relativity (GR), research interests in the geometric gravity theory of Newton-Cartan remains steady due to it being a potential breakthrough for a working quantum theory of gravity.  The Horava-Lifschitz gravity theory is an example of Newton-Cartan application\cite{Horava}; such theories obtains candidacy position on quantum gravity theory by dropping the isotropic space and time assumption.  Hence, adopting classical spacetime structure and separating space and time symmetries, Newton-Cartan is indeed compatible with quantum symmetries such as the conformal symmetries of the Schrodinger's algebra\cite{Schrodinger1,Schrodinger2}.  

\paragraph{}
In our previous work, we have derived and discussed the Newton-Cartan geometric structure from the gauging of Bargmann algebra\cite{Bargmann} - which is obtained by performing Inonu-Wigner $c\rightarrow\infty$ contraction\cite{Inonu} of the Poincaré algebra - as inspired by previous literatures such as Ref.\cite{Bergshoeff,HartongObers}.  Despite being a fundamental and rigorous method capable of deriving solutions of classical connections $\Gamma^\lambda_{\mu \nu}$, the general "line element" structure is not immediately obvious from the gauging process: as we shall see a classical line element are not always metric determined!  On this aspect, the direct metric $1/c^2$ expansion method of Ref.\cite{Bleeken} is more applicable.  Rather expectedly, Ref.\cite{Bleeken} confirms that Newton-Cartan structure emerges as one directly expands GR's metric and $\Gamma^\lambda_{\mu \nu}$ up to order $\mathcal{O}(1/c^2)$.  

\paragraph{}
From these results, the Newton-Cartan limit of any given metrics can in principal be derived and the corresponding classical analogue equation of motions can be obtained.  In this paper, we shall do so for the geometry of a Schwarzschild spacetime as its Newtonian limit can be highly relevant to our everyday gravity.  Interestingly, Ref.\cite{Bleeken}'s results imply that a $c\rightarrow\infty$ limit of (the non-torsional) relativistic Schwarzschild geometry might yield both non-torsional Newton-Cartan geometry (TLNC) and twistless torsional Newton-Cartan geometry (TTNC), depending on our assumption of the gravitational field strength.  TTNC has recently gained more research interests\cite{Hartong2,Bergshoeff2,Bekaert,Banerjee} due to it being a feasible generalization of its non-torsional counterpart and due to its "twistless" condition being an interesting physical manifestation of the Frobenius condition\cite{Frobenius}.  

\paragraph{}
The explicit expressions of both non-torsional and torsional classical metric analogue of a well known GR geometry should certainly be of high interest: it allows one concretely analyze the fundamental differences between the three theories.  These include the differing temporal structures, the manifestation of torsion, the geometry implications on the rules of parallel transport and geodesics, etc.  In this paper, we shall attempt this analysis by borrowing a method from gauge theory: the path integral formulation and holonomy map.  We shall elaborate the reasonings of our choice in the discussion; in a nutshell, such formulation is more fundamental/universal, more compact and is in general simpler to compute rather than deriving the same results by solving the transport/geodesic equations one by one.  To compute the path integral quantities, one simply needs two components: the connections coefficient\footnote{Either the linear connections $\Gamma^\lambda_{\mu \nu}$ or the vielbein connections $\omega^{AB}_\mu$ which is more general.} and the generators corresponding to the gauge field.  The gauge field in this case is nothing but the general Lie algebra valued one forms of the corresponding symmetry group.  The resulting path integral would then determine particle trajectories, vector evolutions under transport, etc.  The holonomy map on the other hand refers to the path integral quantity when the path is a closed loop: this gives us Lie algebra valued elements that we refer to as the holonomy groups.  

\paragraph{}
Holonomy groups are perhaps the most compact quantity capable of describing a manifold's characteristics as the information of curvatures, torsions etc. are encoded inside it.  Furthermore, the application of this notion on the spacetime manifold has received considerable interests in previous works and literatures e.g. Ref.\cite{Gambini,Anandan}, mainly due to its uncanny (yet expected, as we shall discuss) resemblance to geometrical phases in quantum mechanics and the potential application to path integral quantization of gravity. With this in mind and the fact that Newton-Cartan manifolds do accommodate the quantum framework, the idea to form quantum-gravity connection by studying Newton-Cartan holonomy groups might not be too far-fetched after all.


\section{Methodology} \label{sec2}

\paragraph{}
Throughout this work, we shall use the results and formulations found in the Newton-Cartan theory (see, for example \cite{Lian, HartongObers, Bergshoeff}).  From the point of view of gauge theory, Newton-Cartan differs from GR in terms of their total gauge field $A_\mu$.  While $A_\mu$ for the latter is extracted from the Poincaré symmetry group\footnote{The convention throughout this paper is as follows: $\{A, B, C, \dots\}$ denote spacetime vielbein indices, $\{a, b, c, \dots\}$ denote spatial vielbein indices, and $\{\mu, \nu, \sigma, \dots\}$ denote coordinate spacetime indices.  For generators, $P_A$ denotes spacetime translation generator, $M_{AB}$ denotes Lorentz boost generator, $H$ denotes the Hamiltonian/ time translation generator, $P_a$ denotes spatial translation generator, $G_a$ denotes Galilean boost generator, $J_{ab}$ denotes spatial rotation generator, and $M$ denotes the central extension/mass generator.}

\begin{equation} \label{eq1}
A_\mu = P_A e^A_\mu + \frac{1}{2} M_{AB} \omega^{AB}_\mu,
\end{equation}

the Bargmann symmetry group\cite{Bargmann} - which is a central extension of the Galilean group - underlies the former's:\footnote{Strictly speaking, the translation parts of both Poincare and Bargmann algebra must be replaced by a diffeomorphism transform for them to be the symmetry group, as discussed in Ref.\cite{Lian}.}

\begin{equation} \label{eq2}
A_\mu= H \tau_\mu + P_a e^a_\mu + G_a \omega^a_\mu + \frac{1}{2} J_{ab} \omega^{ab}_\mu + M m_\mu,
\end{equation}

\paragraph{}
As was done in previous works \cite{HartongObers, Bergshoeff} - due to their matching transformation properties under the corresponding Lie algebra elements\cite{Lian} - we have used the freedom to identify the above translational and boost/rotational gauge fields as the vielbeins and the vielbein connections, respectively.  

\paragraph{}
While in GR and its corresponding Poincaré algebra one is introduced to a set of spacetime vielbeins $e^A_\mu$, in Bargmann algebra such set is represented by the union of a set of spatial vielbeins $e^a_\mu$, a temporal vielbein $\tau_\mu$ and the vielbein generated by the mass gauge field $m_\mu$.  Using these vielbeins, one creates the boost/rotation invariant quantities which will in turn be used to construct invariant spacetime parameters such as linear connections.  The actualization of the last statement is straightforward in GR: the metric tensor $g_{\mu \nu}=\delta_{ab} e^a_\mu e^b_\nu$ - as well as $g^{\mu \nu}=\delta^{ab} e^\mu_a e^\nu_b$ - are natural invariants and as such able to construct $\Gamma^\lambda_{\mu \nu}$ directly.  This is not true in the Newton-Cartan scheme: invariant quanitites are only obtainable by combining the traditional spatial and temporal vielbeins with $m_\mu$.  As a consequence, Newton-Cartan solutions of $\Gamma^\lambda_{\mu \nu}$ are in general $m_\mu$ dependent.

\paragraph{}
In this work, the solution class of Newton-Cartan $\Gamma^\lambda_{\mu \nu}$ that is linear to $m_\mu$ shall be adopted:

\begin{equation} \label{eq3}
\Gamma^\lambda_{\mu \nu}=-v^\lambda \partial_\mu \tau_\nu + \frac{1}{2} h^{\lambda \sigma} (\partial_\mu h_{\nu \sigma} +\partial_\nu h_{\mu \sigma} -\partial_\sigma h_{\mu \nu}) + \frac{1}{2} h^{\lambda \sigma } (L_{\sigma \mu \nu} + \tau_\mu K_{\sigma \nu} + \tau_\nu K_{\sigma \mu}),
\end{equation}

where 

\begin{align}
K_{\mu \nu}&= 2 \partial_{[\mu} m_{\nu]}       \label{eq4}\\
L_{\sigma \mu \nu}&=  2m_\sigma \partial_{[\mu} \tau_{\nu]} -2m_\mu \partial_{[\nu} \tau_{\sigma]} - 2m_\nu \partial_{[\mu} \tau_{\sigma]}                 \label{eq5}
\end{align}

are the Hartong-Obers parameters\cite{HartongObers}.  Indeed, in \ref{sec2.2} it shall be verified that such solution class accommodates the appropriate Newton-Cartan limit of the Schwarzschild linear connection.  

\paragraph{}
In our previous work, we analyzed the physical implications of Eq.\ref{eq3} such as the resulting spatial contortion $\kappa_{\mu \nu \sigma}=-(2\Gamma^\lambda_{[\mu \sigma]} h_{\nu \lambda} +2 \Gamma^\lambda_{[\nu \sigma]} h_{\mu \lambda} -2 \Gamma^\lambda_{[\mu \nu]} h_{\lambda \sigma} )$ and Newton-Coriolis two form\footnote{This is the most common nickname for the quantity, used by various recent literatures on Newton-Cartan.} $n_{\mu \nu}=2\omega_{[\mu} e^a_{\nu]}$.  These follows from the natural parameterizations of $\Gamma^\lambda_{\mu \nu}$\cite{Lian}:

\begin{equation} \label{eq6}
\Gamma^\lambda_{\mu \nu}=-v^\lambda \partial_\mu \tau_\nu + \frac{1}{2} h^{\lambda \sigma} (\partial_\mu h_{\nu \sigma} +\partial_\nu h_{\mu \sigma} -\partial_\sigma h_{\mu \nu}) + \frac{1}{2} h^{\lambda \sigma } (\kappa_{\mu \nu \sigma} + \tau_\mu n_{\sigma \nu} + \tau_\nu n_{\sigma \mu}),
\end{equation}

where a quick matching with Eq.\ref{eq3}, \ref{eq4}, and \ref{eq5} yields 

\begin{align}
\kappa_{\mu \nu \sigma}&=     2m_\sigma \partial_{[\mu} \tau_{\nu]} -2  e^a_\mu e^\rho_a m_\rho \partial_{[\nu} \tau_{\sigma]} - 2  e^a_\nu e^\rho m_\rho \partial_{[\mu} \tau_{\sigma]}             \label{eq7}\\
n_{\mu \nu}&=  2 v^\rho m_\rho  \partial_{[\mu} m_{\sigma]}           .              \label{eq8}
\end{align}

On the other hand, the vielbein connections are related to $\Gamma^\lambda_{\mu \nu}$ via 

\begin{equation} \label{eq9}
\omega^{AB}_\mu = -e^A_\lambda e^{B \nu} \Gamma^\lambda_{\mu \nu} + e^{B \nu} \partial_\mu e^A_\nu
\end{equation}
 
in GR, and 

\begin{align} 
\omega^{ab}_\mu &= -e^a_\lambda e^{b \nu} \Gamma^\lambda_{\mu \nu} + e^{b \nu} \partial_\mu e^a_\nu    \label{eq10}    \\ 
\omega^a_\mu &=   -e^a_\lambda v^\nu \Gamma^\lambda_{\mu \nu} + v^\nu \partial_\mu e^a_\nu        .\label{eq11}
\end{align}

in Newton-Cartan.  Substituting the Levi-Civita connection on Eq.\ref{eq9} and either Newton-Cartan parameterizations of $\Gamma^\lambda_{\mu \nu}$ on Eq.\ref{eq10} and \ref{eq11} yields the following:

\begin{align} 
\omega^{AB}_\mu =&  -e^{A \sigma} e^{B \nu}  \partial_{[\nu} g_{|\mu|\sigma]} +   e^{[B \nu} \partial_\mu e^{A]}_\nu  \label{eq12} \\   
\omega^{ab}_\mu =& -e^{a \sigma} e^{b \nu} ( \partial_{[\nu} h_{|\mu|\sigma]} - m_\sigma \partial_{[\mu} \tau_{\nu]} +m_\mu \partial_{[\nu} \tau_{\sigma]} + m_\nu \partial_{[\mu} \tau_{\sigma]} +\tau_\mu \partial_{[\nu} m_{\sigma]} ) \nonumber \\
&+   e^{[b \nu} \partial_\mu e^{a]}_\nu  \label{eq13} \\   
\omega^a_\mu =&   -e^{a \sigma} v^{\nu} ( \partial_{[\nu} h_{|\mu|\sigma]} - m_\sigma \partial_{[\mu} \tau_{\nu]} +m_\mu \partial_{[\nu} \tau_{\sigma]} + m_\nu \partial_{[\mu} \tau_{\sigma]} ) \nonumber \\  
&+  e^{a \sigma} (\tau_\mu v^\nu \partial_{[\nu} m_{\sigma]} - \partial_{[\mu} m_{\sigma]})  + \frac{1}{2} v^\nu \partial_\mu e^a_\nu      .\label{eq14}
\end{align}

\paragraph{}
Equations of motion and hence holonomies are encoded in Eq.\ref{eq12}, \ref{eq13}, and \ref{eq14}; their information includes the spacetime-vielbein curvature combination ($\partial_\sigma g_{\mu \nu}$ or $\partial_\sigma h_{\mu \nu}$ and  $\partial_\mu e^A_\nu$ or $\partial_\mu e^a_\nu$ terms), spacetime torsions ($\partial_{[\mu} \tau_{\nu]}$ terms), and Coriolis forces ($\partial_{[\mu} m_{\nu]}$ terms).

\paragraph{}
In the context of Newton-Cartan, it is interesting to study how Eq.\ref{eq13} and \ref{eq14} reform under various physical conditions.  In this work, we are interested on analyzing two physical spacetime structure: the torsionless Newton-Cartan geometry (TLNC) where $d\mathbf{\tau}=0$ and the twistless torsional Newton-Cartan geometry (TTNC) which is a torsional spacetime with the Frobenius condition\cite{Frobenius} $\mathbf{\tau} \land d\mathbf{\tau}=0$ on its time hypersurfaces.  

\paragraph{}
It is straightforward to see that the zero temporal torsion of TLNC suffices to ensure zero spatial contortion $\kappa_{\mu \nu \sigma}$ and consequently allows a simpler solutions for $\omega^{ab}_\mu$ and $\omega^a_\mu$.

\begin{align} 
\overset{\mathrm{TLNC}}{\omega^{ab}_\mu} &= -e^{a \sigma} e^{b \nu} ( \partial_{[\nu} h_{|\mu|\sigma]}  +\tau_\mu \partial_{[\nu} m_{\sigma]} )+   e^{[b \nu} \partial_\mu e^{a]}_\nu   \label{eq15} \\   
\overset{\mathrm{TLNC}}{\omega^a_\mu} &=     -e^{a \sigma} v^{\nu}  \partial_{[\nu} h_{|\mu|\sigma]}+  e^{a \sigma} (\tau_\mu v^\nu \partial_{[\nu} m_{\sigma]} - \partial_{[\mu} m_{\sigma]})  + \frac{1}{2} v^\nu \partial_\mu e^a_\nu      .\label{eq16}
\end{align}

\paragraph{}
Finally, while demanding $m_\mu=0$ for every observer is not possible in Newton-Cartan formalism ($m_\mu$ is required to balance the non-invariance of other vielbeins), certain metrics allow vanishing $m_\mu$ for certain observers.  It is interesting to note that for such observers, quantities in Eq.\ref{eq7} and \ref{eq8} also vanish implying no torsion nor Coriolis forces are observable despite their actual global existence.  In this case, Eq.\ref{eq13} and \ref{eq14} reduces significantly to 

\begin{align} 
\overset{m_\mu=0}{\omega^{ab}_\mu} &= -e^{a \sigma} e^{b \nu}  \partial_{[\nu} h_{|\mu|\sigma]} +  e^{[b \nu} \partial_\mu e^{a]}_\nu    \label{eq17} \\   
\overset{m_\mu=0}{\omega^a_\mu} &=   -e^{a \sigma} v^{\nu}  \partial_{[\nu} h_{|\mu|\sigma]}+ \frac{1}{2} v^\nu \partial_\mu e^a_\nu         .\label{eq18}
\end{align}

which resembles GR's Eq.\ref{eq12} as only the Levi-Civita part of $\Gamma^\lambda_{\mu \nu}$ is left.

\subsection{Path integral formulation} \label{sec2.1}

In the vielbein bundle formalism (see, for example \cite{Isham}), the parallel transport of a vector on a real spacetime base manifold $M$ can be formulated via horizontal lift of the corresponding integral curve onto the bundle space $P$.  More precisely, given a curve in the spacetime $\alpha(s)$, the section function $\sigma : M \rightarrow P$, and the underlying Lie group $G$ representation whose right action is defined on $P$, the corresponding curve of the horizontally lifted vielbein section $\alpha^\uparrow(s)$ is given by

\begin{align}
\alpha^\uparrow(s)&= \sigma(\alpha(s)) g(s) \hspace{0.3cm} \text{where}\nonumber \\
g(s) \in G \hspace{0.3cm} \text{is determined by} \hspace{0.3cm} &\frac{dg(s)}{ds}= -A_\mu(\alpha(s)) \frac{dx^\mu}{ds} (\alpha(s)) g(s)    \label {eq19}
\end{align}

\paragraph{}
and $A_\mu$ the corresponding local connection on the bundle.  Representing $G$ as matrix group, Eq.\ref{eq19} is a principally solvable matrix differential equation

\begin{align}
g(s)&=g_0 -\int A_\mu \frac{dx^\mu}{ds} g(s')   ds'    \hspace{0.3cm}\text{, or} \label {eq20}\\ 
g(s)&=g_0 \mathcal{P} \text{exp} \Big( -\int A_\mu \frac{dx^\mu}{ds'}  ds' \Big)  \label {eq21}
\end{align} 

where $\mathcal{P}$ denotes path ordering.  The closed loop holonomy map is simply $g(s)$ corresponding to a closed loop integral.

\paragraph{}
Our convention and interpretation of the system is as follows: as the horizontal vielbein section, $\alpha^\uparrow(s)$ contains an observer's horizontally evolving frame set $[\tilde{e}^A_\mu(s)]$ or $[\tilde{e}^a_\mu(s), \tilde{\tau}_\mu(s)]$ that is required to be orthonormal everywhere.  At $s=0$, we set $\alpha^\uparrow(0)$ to be the natural orthonormal frame at the origin; the natural frame are nothing but the metric vielbeins themselves i.e. $e^a_\mu$ that were extracted from $g_{\mu \nu}$.  $\sigma(\alpha(s))$ on the other hand always represent the natural frame at all points.  With this setting, $g_0$ is required to be the identity, and a general $g(s)$ transforms a natural frame onto the correct horizontal frame of the transported observer.  Note that $g(s)$'s domain and range are restricted to orthonormal frame sections.  

\paragraph{}
Should the evolution of the transported vector $X=X^\mu \frac{\partial}{\partial x^\mu}$ is of interest, it can be read from $g(s)$ as follows: 

\begin{equation}
X^a (s) = X^\mu (s) e^a_\mu (s) = X^\mu \tilde{e}^a_\mu (s) = X^\mu \sigma(\alpha(s)) g(s) \label{eq22}
\end{equation}

where $X^a (s)$ ($X^\mu (s)$) is the evolving vector's $a$ th ($\mu$ th) vielbein (coordinate) component.\footnote{Note that $X^\mu$ denotes a pre-determined vector field/ integral curve of $X$ whereas $X^\mu(s)$ denotes how the observer measures $X$ in its evolving frame $\tilde{e}^a_\mu (s)$.}

\paragraph{}
Before proceeding to compute $A_\mu$ in each case, it is important to note that $A_\mu$ in Eq.\ref{eq19}, \ref{eq20} and \ref{eq21} are not given by Eq.\ref{eq1} and \ref{eq2}.  In the context of vielbein bundles, $A_\mu$ is a $G$-valued horizontal connection: its action on a section on a bundle can only be a vertical displacement along the fiber and not horizontal to another fiber belonging to another point in the spacetime.  The inclusion of $P_A$, $H$, and $P_a$ in $A_\mu$ violates this condition\footnote{Once again, as discussed in Ref.\cite{Lian}, this is a consequence of translations not being an internal symmetry of these geometric theories of gravity.  The non-modified Poincare or Bargmann algebra are not symmetry groups by themselves, but their rotation/boosts subgroups are.} since they induce displacements on the base manifold.  As a consequence, we shall use the general $G$-valued horizontal connection as $A_\mu$:

\begin{equation} \label{eq23}
A_\mu = \frac{1}{2} M_{AB} \omega^{AB}_\mu \hspace{0.3cm} \text{in GR}
\end{equation}

\begin{equation} \label{eq24}
A_\mu= G_a \omega^a_\mu + \frac{1}{2} J_{ab} \omega^{ab}_\mu \hspace{0.3cm} \text{in Newton-Cartan}
\end{equation}

\paragraph{}
To verify the above, consider the case of a flat spacetime without gravity and where a trivial/coordinate vielbeins are used.  In such setting, the obvious solution for $g(s)$ is the identity matrix at any $s$ since vectors will remain unchanged along any path.  Non-trivial $g(s)$'s would nevertheless be obtained if one uses Eq.\ref{eq1} and \ref{eq2} as $A_\mu$, contradicting the obvious conclusion.  There is simply no reason for the transported vector to be arbitrarily displaced to another point in time, space, or state of $m_\mu$ along its path based on an arbitrary choice of vielbeins as their gauge field.  Under Eq.\ref{eq23} and \ref{eq24}, vectors merely rotates and boosts due to the spacetime curvature, torsion and Coriolis force encoded in the vielbein connections as they should be.

\subsection{Relativistic Schwarzschild connections}   \label{sec2.2}

The relativistic Schwarzschild line element is given by the familiar

\begin{equation} \label{eq25}
ds^2= -\big(1-\frac{2M}{rc^2}\big) c^2 dt^2 + \big(1-\frac{2M}{rc^2}\big)^{-1} dr^2 + r^2 \sin^2\theta d\theta^2 + r^2 d\phi^2,
\end{equation} 

from which the vielbeins i.e. the natural frames can be extracted:\footnote{Notice that throughout this paper, our coordinate is $(t, r, \theta, \phi)$ instead of $(ct, r, \theta, \phi)$ to accommodate comparisons with normal classical coordinates.}	

\begin{equation}\label{eq26}
e^A_\mu = [e^1_\mu, e^2_\mu, e^3_\mu, e^4_\mu] =
\begin{bmatrix} \big(1-\frac{2M}{rc^2}\big)^{-\frac{1}{2}} &0&0&0\\0&r \sin \theta &0 &0 \\ 0& 0& r &0 \\ 0&0&0&\big(1-\frac{2M}{rc^2}\big)^{\frac{1}{2}}c\end{bmatrix}.
\end{equation}

Using Eq.\ref{eq12}, the non-zero $\omega^{AB}_\mu$ are

\begin{equation} \label{eq27}
\omega^{12}_\theta = \big(1-\frac{2M}{rc^2}\big)^\frac{1}{2}, \hspace{0.3cm} \omega^{13}_\phi = \big(1-\frac{2M}{rc^2}\big)^\frac{1}{2} \sin \theta, \hspace{0.3cm} \omega^{23}_\phi = \cos \theta, \hspace{0.3cm} \omega^{01}_t = \frac{M}{r^2c}
\end{equation}

and their antisymmetric counterparts obtained by sign flips.  

\subsection{Newton-Cartan Schwarzschild metric and connections}   \label{sec2.3}

\paragraph{}
The Newton-Cartan Schwarzschild structure is obtained by taking the $c\rightarrow \infty$ limit of Eq.\ref{eq25} as worked by Ref.\cite{Bleeken}; in fact, this study shows both TLNC and TTNC geometry can be obtained by adopting different assumptions on the gravitational field strength relative to $c^2$ and preserving only terms of order not lower than $\mathcal{O}(1)$.

\paragraph{}
The TLNC Schwarzschild assumes the field strength $M$ to be of lower order than $c^2$ so that the $\frac{M}{r c^2}$ of $(cdt)^2$ has order $\mathcal{O}(1)$ while the $\frac{M}{r c^2}$ of $dr^2$ has order $\mathcal{O}(1/c^2)$ and is therefore omitted. 

\begin{equation}
ds^2= -  c^2 dt^2 + \frac{2M}{r} dt^2+ dr^2 + r^2 \sin^2\theta d\theta^2 + r^2 d\phi^2, \label{eq28}
\end{equation} 

\paragraph{}
The TLNC separate temporal and spatial metrics can then be read off from the coefficients of $dt^2$, $dr^2$, $d\theta^2$, and $d\phi^2$ (excluding the $\frac{2M}{r} dt^2$ as it is of different order than the $c^2 dt^2$) as follows:

\begin{equation} \label{eq29}
\tau_{\mu \nu}= \begin{bmatrix} 0&0&0&0\\0&0&0&0\\0&0&0&0\\0&0&0&-c^2  \end{bmatrix}, \hspace{0.3cm}
h_{\mu \nu} = \begin{bmatrix} 1&0&0&0\\0&r^2 \sin^2 \theta&0&0\\0&0&r^2&0\\0&0&0&0  \end{bmatrix}.
\end{equation}

The $\frac{2M}{r} dt^2$ term is on the other hand represented by the field\footnote{$m_\mu$ in this case evidently reflects the Newtonian potential; the physical interpretation of $m_0$ appears to be a description of the observer's potential energy as function of spacetime.  The emergence of such classical potential is of course expected in this $c\rightarrow \infty$ limit; it is nevertheless interesting to note that its origin is no other than the relativistic time warp/dilation factor.}

\begin{equation} \label{eq30}
m_{\mu} = \begin{bmatrix} 0&0&0&0\\0&0&0&0\\0&0&0&0\\0&0&0&-\frac{M}{rc}  \end{bmatrix}.
\end{equation}

The torsionless condition is evident from the constant $\tau$.  As such, in our computation Eq.\ref{eq15} and \ref{eq16} shall be used.  The vielbeins or natural frames in this case are given by 

\begin{equation} \label{eq31}
e^a_\mu = [e^1_\mu, e^2_\mu, e^3_\mu] =
\begin{bmatrix} 1&0&0\\0&r \sin \theta&0\\0&0&r\\0&0&0\end{bmatrix}, \hspace{0.3cm}
\tau_\mu=\begin{bmatrix} 0\\0\\0\\c\end{bmatrix}.
\end{equation}

This yields the following non-zero $\omega^{ab}_\mu$ and $\omega^a_\mu$:

\begin{equation} \label{eq32}
\omega^{12}_\theta = 1, \hspace{0.3cm} \omega^{13}_\phi =\sin \theta, \hspace{0.3cm} \omega^{23}_\phi = \cos \theta, \hspace{0.3cm} \omega^{1}_t = \frac{M}{r^2c}
\end{equation}

and the antisymmetric counterparts of the $\omega^{ab}_\mu$ obtained by sign flips.  

\paragraph{}
The TTNC case on the other hand assumes $\frac{M}{r c^2}$ to be of order $\mathcal{O}(1)$ i.e. a strong field condition.  In this case, the line element is still given by Eq.\ref{eq25} but the metrics are nevertheless degenerate in structure.

\begin{equation} \label{eq33}
\tau_{\mu \nu}= \begin{bmatrix} 0&0&0&0\\0&0&0&0\\0&0&0&0\\0&0&0&-c^2  \end{bmatrix}, \hspace{0.3cm}
h_{\mu \nu} = \begin{bmatrix} 1&0&0&0\\0&r^2 \sin^2 \theta&0&0\\0&0&r^2&0\\0&0&0&0  \end{bmatrix}.
\end{equation}
 
The $m_\mu$ field in this case vanishes and the simplified Eq.\ref{eq17} and \ref{eq18} may be used.  The natural frames in TTNC are

\begin{equation} \label{eq34}
e^a_\mu = [e^1_\mu, e^2_\mu, e^3_\mu] =
\begin{bmatrix} \big(1-\frac{2M}{rc^2}\big)^{-\frac{1}{2}} &0&0\\0&r \sin \theta &0 \\ 0& 0& r \\ 0&0&0\end{bmatrix}, \hspace{0.3cm}
\tau_\mu=\begin{bmatrix} 0\\0\\0\\\big(1-\frac{2M}{rc^2}\big)^{\frac{1}{2}}c \end{bmatrix},
\end{equation}

and hence the non-zero $\omega^{ab}_\mu$  are

\begin{equation} \label{eq35}
\omega^{12}_\theta = \big(1-\frac{2M}{rc^2}\big)^\frac{1}{2}, \hspace{0.3cm} \omega^{13}_\phi = \big(1-\frac{2M}{rc^2}\big)^\frac{1}{2} \sin \theta, \hspace{0.3cm} \omega^{23}_\phi = \cos \theta
\end{equation}

and their antisymmetric counterparts obtained by sign flips.  All $\omega^a_\mu$ vanish in this case.  Finally, note that the non-(anti)symmetry of $\partial_r \tau_t=\frac{M}{r^2c}$ and $\partial_t \tau_r=0$ implies the existence of temporal (but not spatial) torsion.


\section{Results} \label{sec3}

\paragraph{}
For each of the above geometries, four types of paths/loops in spacetime shall be analyzed; this is inspired by the methodology of Ref.\cite{Rothman}.  These includes two circular paths: one with evolving time and following the geodesic, the other with frozen time, a radial geodesic loop (observer travelling in geodesic from $r$ to $r'$ and back to $r$), and finally a stationary observer moving only through time (a trivial loop).
\paragraph{}
We shall use the usual transposed\footnote{This is for consistency with our right action convention; our vector-matrix products shall also be $\vv{x} A$ with column $x$ instead of $A\vv{x}$ with row $x$} matrix representation of $J_{ab}$ and $G_a$ throughout e.g. $J^T_{13}=\begin{bmatrix} 0&0&1&0 \\0&0&0&0 \\-1&0&0&0 \\ 0&0 &0 &0\end{bmatrix}$, $G^T_1=\begin{bmatrix} 0&0&0&1 \\0&0&0&0 \\0&0&0&0 \\ 1&0 &0 &0\end{bmatrix}$ and so on. 

\subsection{Holonomy in Relativistic Schwarzschild} \label{sec3.1}

\paragraph{}
To make comparison with Newton-Cartan easier, we shall decompose the Lorentz boost generator $M_{AB}$ into separate rotation and boost generators: $M_{ab}=J_{ab}$ and $M_{0a}= -M_{a0}= G_a$\footnote{$M_{0a}\neq cG_a$ since $M_{AB}$ already acts as generator of $t$, not $ct$} for spatial indices $\{a,b\}$.

\subsubsection{Circular timeless loop}   \label{sec3.1.1}

Without loss of generality, we shall consider $\phi$ circular loop at the equator $\theta=\frac{\pi}{2}$. Eq.\ref{eq19} reads

\begin{equation} \label{eq36}
\frac{dg(\phi)}{d\phi}= -  \big(1-\frac{2M}{rc^2} \big)^\frac{1}{2}  J^T_{13}    g(\phi)
\end{equation}

which - with $g(\phi=0)=I$ - is solvable as

\begin{equation} \label{eq37}
g(\phi)= \mathcal{P} \text{exp} \bigg(  -\big(1-\frac{2M}{rc^2} \big)^\frac{1}{2}  J^T_{13}  \phi \bigg),
\end{equation}

or, 

\begin{equation} \label{eq38}
g(\phi)=\begin{bmatrix} \cos(\Lambda \phi)&0  & -\big(1-\frac{2M}{rc^2} \big)^\frac{1}{2} \frac{1}{\Lambda} \sin (\Lambda \phi ) & 0
\\ 0 &1 &0 &0  
\\ \big(1-\frac{2M}{rc^2} \big)^\frac{1}{2} \frac{1}{\Lambda} \sin (\Lambda \phi ) & 0 & \frac{1}{\Lambda^2} \big(1-\frac{2M}{rc^2} \big) \cos(\Lambda \phi)   & 0
\\ 0  &0 &  0 &  1 \end{bmatrix}.
\end{equation}

where $\Lambda= \big(1-\frac{2M}{rc^2} \big)^\frac{1}{2}$.  The frame evolution $[\tilde{e}^1, \tilde{e}^2, \tilde{e}^3, \tilde{e}^4]=\sigma g$ corresponds to 

\begin{equation} \label{eq39}
\begin{bmatrix} \big(1-\frac{2M}{rc^2} \big)^{-\frac{1}{2}} \cos(\Lambda \phi)&0  & -\frac{1}{\Lambda} \sin(\Lambda \phi) & 0  
\\ 0 &r &0 &0  
\\ r  \sin (\Lambda \phi )& 0 & r \cos(\Lambda \phi) &0
\\ 0 &0 & 0  &  \big(1-\frac{2M}{rc^2} \big)^\frac{1}{2}   \end{bmatrix}.
\end{equation}

\subsubsection{Circular geodesic loop}  \label{sec3.1.2}

With $s=\phi$, $\theta=\frac{\pi}{2}$, and assuming in the geodesic $\frac{dt}{d \phi}=\mu$, Eq.\ref{eq19} in reads 

\begin{equation} \label{eq40}
\frac{dg(\phi)}{d\phi}= - \big[ \big(1-\frac{2M}{rc^2} \big)^\frac{1}{2}  J^T_{13}   + \mu \frac{M}{r^2 c}  G^T_1 \big] g(\phi)
\end{equation}

This is again solvable as 

\begin{equation} \label{eq41}
g(\phi)= \mathcal{P} \text{exp} \bigg( -\Big( \big(1-\frac{2M}{rc^2} \big)^\frac{1}{2}  J^T_{13}  + \mu \frac{M}{r^2 c}  G^T_1  \Big)\phi \bigg),
\end{equation}

but due to non-commutativity of $J^T_{13}$ and $G^T_1$, one needs to apply the Zessenhaus\cite{Zessenhaus} formula to expand $g(\phi)$ and solve the resulting infinite series, which is not a trivial computation.  In this case, it is more efficient to solve the matrix equation column by column:

\begin{equation} \label{eq42}
\frac{d}{d\phi}\begin{bmatrix} g_{1i} \\ g_{2i} \\g_{3i} \\ g_{4i} \end{bmatrix}= \begin{bmatrix} -\frac{1}{r} g_{3i} - \mu \frac{M}{r^2c}   \big(1-\frac{2M}{rc^2} \big)^{-1} g_{4i} \\ 0\\r  \big(1-\frac{2M}{rc^2} \big) g_{1i}  \\ - \big(1-\frac{2M}{rc^2} \big)  \mu \frac{M}{r^2 c}   g_{1i}  \end{bmatrix}.
\end{equation}

The oscillating frequency in this case is given by $\omega= \big(1-\frac{2M}{rc^2} - \mu^2 \frac{M^2}{ r^4 c^2}\big)^\frac{1}{2}$.  With initial condition $g(\phi=0)=I$, one obtains

\begin{widerequation} \label{eq43}
g(\phi)=\begin{bmatrix} \cos(\omega \phi)&0  & -\big(1-\frac{2M}{rc^2} \big)^\frac{1}{2} \frac{1}{\omega} \sin (\omega \phi ) & -\mu \frac{M}{r^2c}  \frac{1}{\omega} \sin (\omega \phi )  
\\ 0 &1 &0 &0  
\\ \big(1-\frac{2M}{rc^2} \big)^\frac{1}{2} \frac{1}{\omega} \sin (\omega \phi ) & 0 & \frac{1}{\omega^2} \big(1-\frac{2M}{rc^2} \big) \cos(\omega \phi) -  \mu^2 \frac{M^2}{\omega^2 r^4 c^2} & \big(1-\frac{2M}{rc^2} \big)^\frac{1}{2}  \mu \frac{M}{r^2c} \frac{1}{\omega^2} (\cos(\omega \phi) -1)
\\ -\mu \frac{M}{r^2c}  \frac{1}{\omega} \sin (\omega \phi )   &0 &  -\big(1-\frac{2M}{rc^2} \big)^\frac{1}{2}  \mu \frac{M}{r^2c} \frac{1}{\omega^2} (\cos(\omega \phi) -1)  &  \frac{1}{\omega^2} \Big(\big(1-\frac{2M}{rc^2} \big) - \mu^2 \frac{M^2}{ r^4 c^2} \cos(\omega \phi) \Big)  \end{bmatrix}.
\end{widerequation}

The observer's frame evolution $[\tilde{e}^1, \tilde{e}^2, \tilde{e}^3, \tilde{e}^4]$ is given by 

\begin{widerequation} \label{eq44}
\begin{bmatrix} \big(1-\frac{2M}{rc^2} \big)^{-\frac{1}{2}} \cos(\omega \phi)&0  & -\frac{1}{\omega} \sin (\omega \phi ) & -\big(1-\frac{2M}{rc^2} \big)^{-\frac{1}{2}}  \mu \frac{M}{r^2c}  \frac{1}{\omega} \sin (\omega \phi )  
\\ 0 &r &0 &0  
\\ r \big(1-\frac{2M}{rc^2} \big)^\frac{1}{2} \frac{1}{\omega} \sin (\omega \phi ) & 0 & \frac{r}{\omega^2} \big(1-\frac{2M}{rc^2} \big) \cos(\omega \phi) -  \mu^2 \frac{M^2}{\omega^2 r^3 c^2} & \big(1-\frac{2M}{rc^2} \big)^\frac{1}{2}  \mu \frac{M}{rc} \frac{1}{\omega} (\cos(\omega \phi) -1)
\\ -\big(1-\frac{2M}{rc^2} \big)^{\frac{1}{2}} \mu \frac{M}{r^2}  \frac{1}{\omega} \sin (\omega \phi )   &0 &  -\big(1-\frac{2M}{rc^2} \big) \mu \frac{M}{r^2} \frac{1}{\omega^2} (\cos(\omega \phi) -1)  &  \frac{c}{\omega^2} \big(1-\frac{2M}{rc^2} \big)^\frac{1}{2}  \Big(\big(1-\frac{2M}{rc^2} \big) - \mu^2 \frac{M^2}{ r^4 c^2} \cos(\omega \phi) \Big)  \end{bmatrix}.
\end{widerequation}

For a circular geodesic, the vector $X=X^\phi \frac{\partial}{\partial \phi} + X^t \frac{\partial}{\partial t}$ transported along $X$'s integral curves must satisfy $r \big(1-\frac{2M}{rc^2} \big)^{\frac{1}{2}} \frac{1}{\omega} \sin (\omega \phi )X^\phi   - \big(1-\frac{2M}{rc^2} \big)^{\frac{1}{2}} \mu \frac{M}{r^2}  \frac{1}{\omega} \sin (\omega \phi )  X^t  =0 $ so that its $X^r$ will remain zero at all times.  Combining this condition with $\frac{X^t}{X^\phi}=\frac{dt}{d\phi}=\mu $ yields the Keppler's law for a circular orbit:

\begin{equation} \label{eq45}
\mu^2 M = r^3.
\end{equation} 

\subsubsection{Radial geodesic loop}    \label{sec3.1.3}

Keeping $\theta$ and $\phi$ constant, and demanding $\frac{dt}{dr}= \frac{1}{c} \big(1-\frac{2M}{rc^2} \big)^{-1}$ to ensure compatibility with the radial geodesic integral curve, Eq.\ref{eq19} for outward direction is given by 

\begin{equation} \label{eq46}
\frac{dg(r)}{dr}= -\big[ \big(1-\frac{2M}{rc^2} \big)^{-1} \frac{M}{r^2c^2}    G^T_1 \big] g(r)
\end{equation}

which can be directly solved as

\begin{equation}\label{eq47}
g(r,r_0)=  \mathcal{P} \text{exp} \bigg( G^T_1 \ln \Big( \big(1-\frac{2M}{rc^2} \big)^{-\frac{1}{2}} /\big(1-\frac{2M}{r_0c^2} \big)^{-\frac{1}{2}}  \Big) \bigg),
\end{equation}

or,

\begin{equation} \label{eq48}
g^{\mathrm{out}}(r,r_0)= \begin{bmatrix} \frac{\beta}{2} + \frac{1}{2\beta} &0 & 0 & \frac{\beta}{2} - \frac{1}{2\beta}\\0&1&0&0 \\0&0&1&0 \\ \frac{\beta}{2} - \frac{1}{2\beta} & 0 &0 &\frac{\beta}{2} + \frac{1}{2\beta} \end{bmatrix},
\end{equation}

$\beta$ being $ \big(1-\frac{2M}{rc^2} \big)^{-\frac{1}{2}}  \big(1-\frac{2M}{r_0c^2} \big)^\frac{1}{2}$.

The inward $g(r)$ can be obtained the same way with $\frac{dt}{dr}= - \frac{1}{c}\big(1-\frac{2M}{r} \big)^{-1}$; this yields

\begin{equation} \label{eq49}
g^{\mathrm{in}}(r,r_0)= \begin{bmatrix} \frac{\beta}{2} + \frac{1}{2\beta} &0 & 0 & -\frac{\beta}{2} + \frac{1}{2\beta}\\0&1&0&0 \\0&0&1&0 \\ -\frac{\beta}{2} + \frac{1}{2\beta} & 0 &0 &\frac{\beta}{2} + \frac{1}{2\beta} \end{bmatrix}.
\end{equation}

When a vector is transported from $r_1$ to $r_2$ and back to $r_1$ with $r_2>r_1$ on these geodesics, the holonomy would be the product of $g^{\mathrm{out}}(r_2,r_1)$ and $g^{\mathrm{in}}(r_1,r_2)$:

\begin{equation}\label{eq50}
\alpha ^{' \uparrow } (r_1) = \sigma(r_1) g^{\mathrm{loop}} (r_1 \rightarrow r_2 \rightarrow r_1) = \sigma(r_1) g^{\mathrm{out}}(r_2,r_1)   g^{\mathrm{in}}(r_1,r_2)
\end{equation}

If we let $\beta_{12}$ to be $ \big(1-\frac{2M}{r_2c^2} \big)^{-\frac{1}{2}}  \big(1-\frac{2M}{r_1c^2} \big)^\frac{1}{2}$, we find 

\begin{equation} \label{eq51}
g^{\mathrm{loop}} (r_1 \rightarrow r_2 \rightarrow r_1)	= \begin{bmatrix} \cosh (2 \ln \beta_{12}) &0 & 0 & \sinh (2 \ln \beta_{12}) \\0&1&0&0 \\0&0&1&0 \\ \sinh (2 \ln \beta_{12}) & 0 &0 &\cosh (2 \ln \beta_{12}) \end{bmatrix}.
\end{equation}

\subsubsection{Stationary loop}            \label{sec3.1.4}

When the observer travels exclusively in time, Eq.\ref{eq19} becomes

\begin{equation} \label{eq52}
\frac{dg(t)}{dt}= -  \frac{M}{r^2c}    G^T_1  g(r),
\end{equation}

that is, 

\begin{equation} \label{eq53}
g(t)= \mathcal{P} \text{exp} \big(  -\frac{M}{r^2c}  G^T_1  t  \big).
\end{equation}

Hence, 

\begin{equation} \label{eq54}
g(t)= \begin{bmatrix} \cosh (\Omega t) &0 & 0 & -\sinh (\Omega t)\\0&1&0&0 \\0&0&1&0 \\ -\sinh (\Omega t) & 0 &0 & \cosh (\Omega t) \end{bmatrix},
\end{equation}

where $\Omega$ is given by $\frac{M}{r^2c}$.

\subsection{Holonomy in TLNC Schwarzschild} \label{sec3.2}

\subsubsection{Circular timeless loop}   \label{sec3.2.1}

It is clear with time not concerned, the TLNC space structure (Eq.\ref{eq29}) is completely flat.  Thus, the timeless circular curve $g(s)$ shall be the same as flat space's $g(s)$ when spherical vielbeins are used. Eq.\ref{eq19} at $\theta=\frac{\pi}{2}$ reads

\begin{equation} \label{eq55}
\frac{dg(\phi)}{d\phi}= - J^T_{13}    g(\phi),
\end{equation}

which can be solved as

\begin{equation} \label{eq56}
g(\phi)= \mathcal{P} \text{exp} \big(  -  J^T_{13}  \phi \big),
\end{equation}

or, 

\begin{equation} \label{eq57}
g(\phi)=\begin{bmatrix} \cos  \phi&0  & -  \sin \phi  & 0
\\ 0 &1 &0 &0  
\\ \sin   \phi  & 0 &  \cos \phi  & 0
\\ 0  &0 &  0 &  1 \end{bmatrix}.
\end{equation}

This implies a trivial holonomy for a closed equatorial $\phi$ loop.\footnote{Note that for $\theta\neq\frac{\pi}{2}$, the holonomy is non-trivial in general and $\frac{dg(\phi)}{d\phi}$ is given by $-(  \sin  \theta J^T_{13} + \cos \theta J^T_{23}    )  g(\phi)$.}

\subsubsection{Circular geodesic loop}    \label{sec3.1.2}

If time is turned on, and assuming $\frac{dt}{d \phi}=\mu$, the equatorial loop geodesic evolution is described by 

\begin{equation} \label{eq58}
\frac{dg(\phi)}{d\phi}= - \big[  J^T_{13}   + \mu \frac{M}{r^2 c}  G^T_1 \big] g(\phi)
\end{equation}

Once again, rather than expanding the exponential solution

\begin{equation} \label{eq59}
g(\phi)= \mathcal{P} \text{exp} \bigg( -\Big(  J^T_{13}  + \mu \frac{M}{r^2 c}  G^T_1  \Big)\phi \bigg)
\end{equation}

manually, we shall instead decompose the matrices onto individual columns: 

\begin{equation}\label{eq60}
\frac{d}{d\phi}\begin{bmatrix} g_{1i} \\ g_{2i} \\g_{3i} \\ g_{4i} \end{bmatrix}= \begin{bmatrix} -\frac{1}{r} g_{3i} - \mu \frac{M}{r^2c}   g_{4i} \\ 0\\r   g_{1i}  \\ -  \mu \frac{M}{r^2 c}   g_{1i}  \end{bmatrix}.
\end{equation}

The oscillating frequency is given by $\omega= \big(1- \mu^2 \frac{M^2}{ r^4 c^2}\big)^\frac{1}{2}$.  With $g(\phi=0)=I$, one obtains

\begin{widerequation} \label{eq61}
g(\phi)=\begin{bmatrix} \cos(\omega \phi)&0  & - \frac{1}{\omega} \sin (\omega \phi ) & -\mu \frac{M}{r^2c}  \frac{1}{\omega} \sin (\omega \phi )  
\\ 0 &1 &0 &0  
\\  \frac{1}{\omega} \sin (\omega \phi ) & 0 & \frac{1}{\omega^2} \cos(\omega \phi) -  \mu^2 \frac{M^2}{\omega^2 r^4 c^2} & \mu \frac{M}{r^2c} \frac{1}{\omega^2} (\cos(\omega \phi) -1)
\\ -\mu \frac{M}{r^2c}  \frac{1}{\omega} \sin (\omega \phi )   &0 &  - \mu \frac{M}{r^2c} \frac{1}{\omega^2} (\cos(\omega \phi) -1)  &  \frac{1}{\omega^2} \Big(1- \mu^2 \frac{M^2}{ r^4 c^2} \cos(\omega \phi) \Big)  \end{bmatrix}.
\end{widerequation}

It is easy to see that the same Keppler's law for circular orbit holds; the observer's frame evolution $[\tilde{e}^1, \tilde{e}^2, \tilde{e}^3, \tilde{e}^4]$ is

\begin{equation} \label{eq62}
\begin{bmatrix} \cos(\omega \phi)&0  & - \frac{1}{\omega} \sin (\omega \phi ) & -\mu \frac{M}{r^2c}  \frac{1}{\omega} \sin (\omega \phi )  
\\ 0 &r &0 &0  
\\  \frac{r}{\omega} \sin (\omega \phi ) & 0 & \frac{r}{\omega^2}  \cos(\omega \phi) -  \mu^2 \frac{M^2}{\omega^2 r^4 c^2} & \mu \frac{M}{rc} \frac{1}{\omega^2} (\cos(\omega \phi) -1)
\\ -\mu \frac{M}{r^2}  \frac{1}{\omega} \sin (\omega \phi )   &0 &  - \mu \frac{M}{r^2} \frac{1}{\omega^2} (\cos(\omega \phi) -1)  &  \frac{c}{\omega^2} \Big(1- \mu^2 \frac{M^2}{ r^4 c^2} \cos(\omega \phi) \Big)  \end{bmatrix}.
\end{equation}

Hence the necessary condition to ensure zero and constant $X^r$ is $ \frac{r}{\omega} \sin (\omega \phi )X^\phi   - \mu \frac{M}{r^2}  \frac{1}{\omega} \sin (\omega \phi )  X^t  =0 $.  Combining this condition with $\frac{X^t}{X^\phi}=\frac{dt}{d\phi}=\mu $ once again yields

\begin{equation} \label{eq63}
\mu^2 M = r^3.
\end{equation}

\subsubsection{Radial geodesic loop}    \label{sec3.1.3}

In TLNC, radial null geodesic is described simply by $\frac{dt}{dr}= \frac{1}{c}$ as there is no warp factors.  Thus, we have

\begin{equation} \label{eq64}
\frac{dg(r)}{dr}= -\frac{M}{r^2c^2}    G^T_1  g(r),
\end{equation}

which solves to

\begin{equation} \label{eq65}
g(r,r_0)=  \mathcal{P} \text{exp} \bigg(  \Big( \frac{M}{r c^2} -  \frac{M}{r_0 c^2}\Big)  G^T_1 \bigg),
\end{equation}

or,

\begin{equation} \label{eq66}
g^{\mathrm{out}}(r,r_0)= \begin{bmatrix} \cosh \gamma &0 & 0 &\sinh \gamma \\0&1&0&0 \\0&0&1&0 \\ \sinh \gamma & 0 &0 &\cosh \gamma \end{bmatrix},
\end{equation}

$\gamma$ being $ \frac{M}{r c^2} -  \frac{M}{r_0 c^2}$.

Similarly, with $\frac{dt}{dr}= - \frac{1}{c}$ the inward geodesic is given by

\begin{equation} \label{eq67}
g^{\mathrm{in}}(r,r_0)= \begin{bmatrix}\cosh \gamma &0 & 0 &-\sinh \gamma \\0&1&0&0 \\0&0&1&0 \\ -\sinh \gamma & 0 &0 &\cosh \gamma \end{bmatrix}.
\end{equation}

The radial loop $r_1 \rightarrow r_2 \rightarrow r_1$ where $r_2>r_1$ is again given by $g^{\mathrm{out}}(r_2,r_1)$ times $g^{\mathrm{in}}(r_1,r_2)$.  Let $\gamma_{12}$ be  $ \frac{M}{r_2 c^2} -  \frac{M}{r_1 c^2}$, then

\begin{equation} \label{eq68}
g^{\mathrm{loop}} (r_1 \rightarrow r_2 \rightarrow r_1)	= \begin{bmatrix} \cosh (2\gamma_{12}) &0 & 0 &\sinh (2\gamma_{12})  \\0&1&0&0 \\0&0&1&0 \\ \sinh(2\gamma_{12})  & 0 &0 &\cosh (2\gamma_{12})  \end{bmatrix}.
\end{equation}

\subsubsection{Stationary loop}            \label{sec3.1.4}

Under stationary condition, TLNC equations are indistinguishable from GR's:

\begin{equation} \label{eq69}
\frac{dg(t)}{dt}= -  \frac{M}{r^2c}    G^T_1  g(r).
\end{equation}

Hence, the same solution as GR solution is obtained

\begin{equation} \label{eq70}
g(t)= \begin{bmatrix} \cosh (\Omega t) &0 & 0 & -\sinh (\Omega t)\\0&1&0&0 \\0&0&1&0 \\ -\sinh (\Omega t) & 0 &0 & \cosh (\Omega t) \end{bmatrix}.
\end{equation}

where $\Omega=\frac{M}{r^2c}$. 

\subsection{Holonomy in TTNC Schwarzschild}  \label{sec3.3}

\subsubsection{Circular timeless loop}   \label{sec3.3.1}

Due to the non-vanishing spatial curvature, TTNC has the same timeless circular holonomy as GR:

\begin{equation} \label{eq71}
\frac{dg(\phi)}{d\phi}= -  \big(1-\frac{2M}{rc^2} \big)^\frac{1}{2}  J^T_{13}    g(\phi),
\end{equation}

whose solution is

\begin{equation} \label{eq72}
g(\phi)=\begin{bmatrix} \cos(\Lambda \phi)&0  & -\big(1-\frac{2M}{rc^2} \big)^\frac{1}{2} \frac{1}{\Lambda} \sin (\Lambda \phi ) & 0
\\ 0 &1 &0 &0  
\\ \big(1-\frac{2M}{rc^2} \big)^\frac{1}{2} \frac{1}{\Lambda} \sin (\Lambda \phi ) & 0 & \frac{1}{\Lambda^2} \big(1-\frac{2M}{rc^2} \big) \cos(\Lambda \phi)   & 0
\\ 0  &0 &  0 &  1 \end{bmatrix}.
\end{equation}

where $\Lambda= \big(1-\frac{2M}{rc^2} \big)^\frac{1}{2}$.

\subsubsection{Circular geodesic loop}    \label{sec3.3.2}

\paragraph{}
Due to the vanishing $\omega^{ab}_t$ and $\omega^a_t$ for any $\{a,b\}$, we find that a circular path in TTNC to be governed by the same equation as it's timeless counterpart.  This means

\begin{equation} \label{eq73}
g(\phi)=\begin{bmatrix} \cos(\omega \phi)&0  & -\big(1-\frac{2M}{rc^2} \big)^\frac{1}{2} \frac{1}{\omega} \sin (\omega \phi ) & 0
\\ 0 &1 &0 &0  
\\ \big(1-\frac{2M}{rc^2} \big)^\frac{1}{2} \frac{1}{\omega} \sin (\omega \phi ) & 0 & \frac{1}{\omega^2} \big(1-\frac{2M}{rc^2} \big) \cos(\omega \phi)   & 0
\\ 0  &0 &  0 &  1 \end{bmatrix}.
\end{equation}

This however is not a geodesic: there exist no value for $X^\phi$ and $X^t$ shall be able to maintain the same ratio throughout the circular motion.  $X^r (\phi)$ component will inevitably emerge if $X^\phi$ is non zero: this implies that any geodesic that includes motion in $\phi$ must also involve motion in $r$.  Thus, a closed loop circular geodesic is non-existent in TTNC. 

\paragraph{}
$\phi$ "orbital" geodesic in this case is in fact described by $ \big(1-\frac{2M}{rc^2} \big)^\frac{1}{2}  cdt = \big(1-\frac{2M}{rc^2} \big)^{-\frac{1}{2}} dr  + r d\phi $; $r$ is a function of $\phi$ and $g(\phi$) would demand a spiral trajectory for geodesic.

\subsubsection{Radial geodesic loop}    \label{sec3.3.3}
\paragraph{}
Another bizzarre consequence of the vanishing $\omega^{ab}_t$ and $\omega^a_t$ is that under radial motion (not only geodesics), one has zero variation for $g$:

\begin{equation} \label{eq74}
\frac{dg(r)}{dr}= 0,\hspace{0.3cm}\text{that is,} \hspace{0.3cm} g(r)=I.
\end{equation}

The observer frame $[\tilde{e}^1, \tilde{e}^2, \tilde{e}^3, \tilde{\tau}]$ as function of $r$ is simply the standard natural vielbein function

\begin{equation} \label{eq75}
\begin{bmatrix} \big(1-\frac{2M}{rc^2}\big)^{-\frac{1}{2}} &0&0&0\\0&r \sin \theta &0 &0\\ 0& 0& r&0 \\ 0&0&0&\big(1-\frac{2M}{rc^2}\big)^\frac{1}{2}\end{bmatrix}.
\end{equation}

Observer's vector under radial motion would be contracted and dilated, but never rotated or boosted in any ways.  Clearly, any radial loop always yields trivial holonomy.

\subsubsection{Stationary loop}            \label{sec3.3.4}
As in the radial motion, $g(t)$ is trivial under any temporal motion:

\begin{equation} \label{eq76}
\frac{dg(t)}{dt}= 0,\hspace{0.3cm}\text{that is,} \hspace{0.3cm} g(t)=I.
\end{equation}

This means transported vectors would remain constant at any point in time.


\section{Discussion} \label{sec4}

\paragraph{}
Several reasonings underlie our preference to represent holonomy as Eq.\ref{eq21} in place of, say, extracting the holonomy from the resulting vector evolution under individual parallel transport equations, or formulating the holonomy in terms of the coordinate bases i.e. $\mathcal{P} \text{exp} \big( -\int  \Gamma_\mu \frac{dx^\mu}{ds} ds \big)$.  The practical reason is because finding $\omega^{AB}_\mu$, $\omega^{ab}_\mu$ and $\omega^a_\mu$ for $A_\mu$ typically requires less effort than finding $\Gamma^\lambda_{\mu \nu}$ in a general torsional spacetime.  It is also more economical to solve one matrix equation than a set of coupled parallel transport differential equations.  Another advantage of Eq.\ref{eq21} is its perceptible expression of physical contribution of each of $A_\mu$'s elemets.  For instance, the element $\omega^{12}_x J_{12}$ of $A_x$ reflects that a non-zero $\omega^{12}_x$ would result in vector rotation from vielbein axis 1 to axis 2 under a displacement on the $x$ coordinate.  

\paragraph{}
Furthermore, path integral formulation of $A_\mu$ interpreted as the gauge field is fundamental and universal: they are nothing but the holonomy map in general gauge theory. If $A_\mu$ has not been specified (i.e. the bundle is not specified to be the vielbein bundle), Eq.\ref{eq20} or \ref{eq21} are applicable to many physical systems.  Substituting electromagnetic potentials for $A_\mu$ yields $g(s)$ that represents the Dirac phase factor in the Aharonov-Bohm effect\cite{AB}, substituting the Hamiltonian for $A_\mu$ yields the solution of the Schrodinger's equation and so on.  As noted by Ref.\cite{Rothman}, the fact that generators such as the Hamiltonian, momentum and rotation operator operates both in real spacetime and the Hilbert space of quantum mechanics\footnote{Again, as noted by \cite{Rothman}, achieving such equivalence still require analytic continuation of the connection to the imaginary axis since the quantum mechanical holonomy formulation differs from Eq.\ref{eq20} and \ref{eq21} by a factor of $i$.} opens up the possibility that such holonomy map is universal in the two realm of physics in the literal sense and there might after all be a relation between parallel transport in spacetime and geometric phases in quantum mechanics.  

\paragraph{}
Going back to our results, table\ref{tab4.1}, \ref{tab4.2}, and \ref{tab4.3} summarize our findings of system evolution/holonomies in the three Schwarzschild geometry scheme.  $\Phi$ in geodesic $r$ refers to the argument of $\cosh$ or $\sinh$ in $g^{\mathrm{loop}} (r_1 \rightarrow r_2 \rightarrow r_1)$.  

\begin{table}[ht]
\centering
\begin{tabular}{|c|c|c|}
\hline
\bf{GR} & $g$ characteristics & Dilation/Contraction\\
\hline 
Timeless $\phi$ loop& sinusoidal, $\Lambda=\big(1-\frac{2M}{rc^2} \big)^\frac{1}{2}$ & Yes  \\
\hline 
Geodesic $\phi$ loop & sinusoidal, $\omega=\big(1-\frac{3M}{rc^2}\big)^\frac{1}{2}$  &  Yes \\
\hline
Geodesic $r$ loop&  hyperbolic, $\Phi=2\ln \beta_{12}$ &  Yes  \\
\hline
Stationary loop& hyperbolic, $\Omega= \frac{M}{r^2c}$  &  Yes  \\
\hline
\end{tabular} 
\caption{Summary of GR Schwarzschild's $g(s)$ and properties}
\label{tab4.1}
\end{table}

\begin{table}[ht]
\centering
\begin{tabular}{|c|c|c|}
\hline
\bf{TLNC} & $g$ characteristics & Dilation/Contraction\\
\hline 
Timeless $\phi$ loop& sinusoidal, $\Lambda=1$ & No  \\
\hline 
Geodesic $\phi$ loop& sinusoidal, $\omega=\big(1-  \frac{M}{r c^2}\big)^\frac{1}{2}$  &  No \\
\hline
Geodesic $r$ loop&  hyperbolic, $\Phi=2 \gamma_{12}$ &  No  \\
\hline
Stationary loop& hyperbolic, $\Omega= \frac{M}{r^2c}$  &  No  \\
\hline
\end{tabular} 
\caption{Summary of TLNC Schwarzschild's $g(s)$ and properties}
\label{tab4.2}
\end{table}

\begin{table}[ht]
\centering
\begin{tabular}{|c|c|c|}
\hline
\bf{TTNC} & $g$ characteristics & Dilation/Contraction\\
\hline 
Timeless $\phi$ loop& sinusoidal, $\Lambda=\big(1-\frac{2M}{rc^2} \big)^\frac{1}{2}$ & Yes  \\
\hline 
Geodesic $\phi$ loop& N/A &  N/A \\
\hline
Geodesic $\phi$ path& spiral &  Yes \\
\hline
Geodesic $r$ loop& trivial  &  Yes \\
\hline
Stationary loop& trivial &  Yes  \\
\hline
\end{tabular} 
\caption{Summary of TTNC Schwarzschild's $g(s)$ and properties}
\label{tab4.3}
\end{table}

\paragraph{}
The key takeaways from these tables are the following:

\begin{enumerate}[label=\arabic*),ref=\arabic*]
\item TLNC is - as expected - partly successful in simulating GR's equation of motions via the $g$ characteristics.  In the two geodesic loop, it yields non-trivial holonomies as GR is, albeit with different sinusoidal frequency or hyperbolic rapidity.  In the case of stationary loop, it even yields the exact GR solution.  Roughly speaking, this means that TLNC is capable to produce sensible equation of motions - resembling GR's form - and provide reliable classical approximation of transport/geodesic equations. 
\item TLNC completely fails to simulate the warp of space and time; no time dilation nor space contraction effect takes place in its $g(s)$'s. 
\item TTNC generally fails to produce sensible equation of motions: its circular geodesic is non-existent and no space-time rotation takes place under radial or temporal motion. This clashes with GR's description of parallel transport.
\item TTNC is completely successful in simulating the effect of time dilation and space contraction.  
\item The holonomy maps/groups can simply be read off the $g$ characteristics as they are simply the full loop $g$'s.  Non-trivial $g$ characteristics generally implies non-trivial holonomy.  Such non-trivialness implies the existence of manifold/vielbein curvature, torsions, or "forces" such as the Coriolis terms. 
\end{enumerate}

\paragraph{}
Our findings of TLNC's characteristics are no surprise: it doesn't admit relativistic notion of spacetime dilation/contraction but as a classical limit theory, it does admit the Newtonian gravitational potential encoded in $m_\mu$.  This $m_\mu$ gives rise to space-time rotations i.e. boosts that in GR was generated by the time dilation effect.  All in all, it can be expected to describe the correct "forces" to describe a particle's motion albeit missing some $\big( 1 -\frac{2M}{rc^2} \big)$ here and there on certain trajectories.

\paragraph{}
Comparisons of parameters such as $\Lambda$, $\omega$, $\Phi$ and $\Omega$ between TLNC and GR are especially practical: they indicate the motions that are affected by dilation/contraction effects only (e.g. timeless $\phi$), or affected by boost effect only (e.g. stationary loop), or affected by both (e.g. the $\phi$ and $r$ geodesic).

\paragraph{}
The case of TTNC is rather tricky and more intriguing. First, being a strong-field theory enforced on a classical background, one might expect it to yield less sensible results.  While this is reflected in its failure to produce sensible equations of motion, we also find its exact description of relativistic dilation/contraction effects.  Of course, such tradeoff is not preferable for a working physical theory as equations of motion is generally of higher priority.

\paragraph{}
TTNC's weakness appears to be its neglection of $m_\mu$ as the Newtonian potential; this eliminates all the necessary boost components of GR.  The $m_\mu$ term that was existent in TLNC was absorbed to be the metric component of $dt^2$, which is the reason for its flawless time dilation description.  While this indeed makes the metric "relativistic", on a classical background where the spatial and temporal metric are separated, this results in the loss of connections between space and time: the $\omega^a_\mu$'s.  To conclude, our result demonstrates the reason a strong-field theory is not compatible with classical structure.  

\paragraph{}
It is nonetheless interesting to analyze the exact mechanism that underlies TTNC's failure.  First, note that by using Eq.\ref{eq22}, one is able to represent the solution in terms of the $X^\mu(s)$'s evolution and thus the transport/geodesic equations in terms of $\Gamma^\lambda_{\mu \nu}$ can be recovered.  This process is straightforward and can be used to confirm our $g(s)$ solutions; in the case of GR, this is done in Appendix A.  In any case, $g(s)$'s behaviour could be understood better by analyzing their corresponding $\Gamma^\lambda_{\mu \nu}$.  

\paragraph{}
For instance, note that Eq.\ref{eq73} on the circular orbit implies 

\begin{align}
X^r (\phi)&=   X^r \cos(\omega \phi) + X^\phi r  \big(1-\frac{2M}{rc^2} \big)^\frac{1}{2} \sin (\omega \phi),  \label{eq77} \\
X^\phi (\phi) &=  -X^r \frac{1}{r\omega} \sin (\omega \phi) + X^\phi  \cos(\omega \phi)     \label{eq78}\\
X^t  (\phi)&= X^t, \label{eq79}
\end{align}

which correspond to GR's transport equations

\begin{align}
\frac{X^r (\phi)}{d\phi} +\Gamma^r_{\phi \phi} X^\phi  (\phi)+ \mu \Gamma^r_{t t} X^t  (\phi) &=   0  \label{eq80} \\
\frac{X^\phi (\phi)}{d\phi} +\Gamma^\phi_{\phi r} X^r (\phi)&=   0 \label{eq81}\\
\frac{X^t (\phi)}{d\phi} + \mu \Gamma^t_{t r} X^r (\phi)&=   0, \label{eq82}
\end{align}

but with $\Gamma^r_{\phi \phi}=-r\sin^2\theta \big(1-\frac{2M}{rc^2} \big)^\frac{1}{2}=-r \big(1-\frac{2M}{rc^2} \big)^\frac{1}{2}$, $\Gamma^r_{t t}=0$, $\Gamma^\phi_{\phi r}=\frac{1}{r}$, and $\Gamma^t_{tr}=0$.

\paragraph{}
Compare these equations with Eq.\ref{eqA7},\ref{eqA8} and \ref{eqA9}: they simply miss the $\Gamma^r_{t t}$ and  $\Gamma^t_{t r}$ terms as these two components vanish in TTNC.  In GR, $\Gamma^t_{t r}$ is simply given by $\Gamma^t_{rt}=\frac{M}{r^2 c}$; this irregularity is of course expected as TTNC has non-vanishing temporal torsion.  The vanishing of $\Gamma^r_{tt}$ on the other hand is a consequence of degenerate metric structure: even though the temporal metric is $r$ dependent and $\partial_r \tau_t\neq0$, such term is not projected onto any spatial direction since time and space are completely separate.   

\paragraph{}
This typical behaviour is again observed in the case of radial loop where our Eq.\ref{eq75} for geodesic initial condition $X^r=c \big(1-\frac{2M}{r_0c^2} \big)^\frac{1}{2}$ and $X^t=\big(1-\frac{2M}{r_0c^2} \big)^{-\frac{1}{2}}$ yields

\begin{align}
X^r (r)&=  c \big(1-\frac{2M}{r_0c^2} \big)^\frac{1}{2}  \label{eq83}\\
X^t  (r)&= \big(1-\frac{2M}{rc^2} \big)^{-\frac{1}{2}}, \label{eq84}
\end{align}

which once again satisfy GR's equations

\begin{align}
\frac{X^r (r)}{dr} + \Gamma^r_{r r} X^r  (r)+  \frac{1}{c} \big(1-\frac{2M}{rc^2} \big)^{-1} \Gamma^r_{t t} X^t  (r) &=   0   \label{eq85}\\
\frac{X^t (r)}{dr} + \frac{1}{c} \big(1-\frac{2M}{rc^2} \big)^{-1} \Gamma^t_{t r} X^r  (r)+   \Gamma^t_{r t} X^t  (r) &=   0,\label{eq86}
\end{align}

but $\Gamma^r_{r r}=\frac{M}{r^2 c} \big(1-\frac{2M}{rc^2} \big)^{-1}$ and $\Gamma^t_{rt}=\frac{M}{r^2 c} \big(1-\frac{2M}{rc^2} \big)$ are the only non-zero $\Gamma^\lambda_{\mu \nu}$.  Comparing with GR, the above equations distinctly lack the  $\Gamma^r_{t t}$ and  $\Gamma^t_{t r}$, just like in the circular geodesic case. 

\paragraph{}
Thus, geometrically speaking TTNC profoundly differs from GR due to its torsional structure and completely separate and degenerate space and time structure.  These prove to be sufficient to deviate fundamental particle trajectories and vector transportation rules.

\paragraph{}
To see why TLNC is much more successful in simulating GR's motion, we note its set of non-zero $\Gamma^\lambda_{\mu \nu}$ corresponding to non-$\theta$ motion: $\Gamma^r_{\phi \phi}=-r$, $\Gamma^r_{t t}=\frac{M}{r^2 c} $, $\Gamma^\phi_{\phi r}=\frac{1}{r}$, and $\Gamma^t_{rt}=\Gamma^t_{tr}=\frac{M}{r^2 c}$, and $\Gamma^r_{r r}=\frac{M}{r^2 c}$.  Albeit missing some $\big( 1 -\frac{2M}{rc^2} \big)$ factors, this set resemble GR's closely in structure.  Not surprisingly, its resulting equations of motion are that of GR's lacking only in dilation/contraction.


\section*{Appendix}

\paragraph{}
We shall demonstrate the equivalence of our results with the usual transport/geodesic equations in the case of GR.  We first expand the results on \ref{sec3.1.2} and \ref{sec3.1.3} in terms of the $X^\mu (s)$ evolutions using Eq.\ref{eq22}.  

\paragraph{}
For the circular geodesic, Eq.\ref{eq44} yields 

\begin{align}
X^r (\phi)=   &X^r \cos(\omega \phi) + X^\phi r \big( 1-\frac{2M}{rc^2} \big) \frac{1}{\omega} \sin (\omega \phi) - X^t  \big( 1-\frac{2M}{rc^2} \big) \mu  \frac{M}{r^2} \frac{1}{\omega} \sin (\omega \phi),  \tag{A1} \label{eqA1}\\
X^\phi (\phi) =  &-X^r \frac{1}{r\omega} \sin (\omega \phi) + X^\phi \Big[ \frac{1}{\omega^2} \big(1-\frac{2M}{rc^2} \big) \cos(\omega \phi) -  \mu^2 \frac{M^2}{\omega^2 r^4 c^2} \Big]  \nonumber \\
& - X^t  \mu \frac{M}{r^3} \big(1-\frac{2M}{rc^2} \big) \frac{1}{\omega^2} (\cos(\omega \phi) -1),    \tag{A2}\label{eqA2}\\
X^t  (\phi)= &-X^r \big( 1-\frac{2M}{rc^2} \big)^{-1} \mu \frac{M}{r^2c}  \frac{1}{\omega} \sin (\omega \phi ) + X^\phi  \mu \frac{M}{rc} \frac{1}{\omega^2} (\cos(\omega \phi) -1) \nonumber \\
&+ X^t \frac{1}{\omega^2} \Big[\big(1-\frac{2M}{rc^2} \big) - \mu^2 \frac{M^2}{ r^4 c^2} \cos(\omega \phi) \Big], \tag{A3}\label{eqA3}
\end{align}

which upon substitution of the geodesic condition $X^\phi=X^t  \mu$ and $X^r=0$ simply becomes 

\begin{align}
X^r (\phi)&=   0  \tag{A4} \label{eqA4}\\
X^\phi (\phi) &=  X^\phi  = X^t (\phi)\mu  \tag{A5}\label{eqA5}\\
X^t  (\phi)&= X^t, \tag{A6}\label{eqA6}
\end{align}

confirming our solution's validity.  It is then straightforward to confirm that Eq.\ref{eqA1}, \ref{eqA2}, and \ref{eqA3} satisfy the transport equations at $\theta=\frac{\pi}{2}$:

\begin{align}
\frac{X^r (\phi)}{d\phi} +\Gamma^r_{\phi \phi} X^\phi  (\phi)+ \mu \Gamma^r_{t t} X^t  (\phi) &=   0   \tag{A7}\label{eqA7}\\
\frac{X^\phi (\phi)}{d\phi} +\Gamma^\phi_{\phi r} X^r (\phi)&=   0 \tag{A8}\label{eqA8}\\
\frac{X^t (\phi)}{d\phi} + \mu \Gamma^t_{t r} X^r (\phi)&=   0, \tag{A9}\label{eqA9}
\end{align}

with $\Gamma^r_{\phi \phi}=-r \sin^2\theta \big(1-\frac{2M}{rc^2} \big)^\frac{1}{2}=-r \big(1-\frac{2M}{rc^2} \big)^\frac{1}{2}$, $\Gamma^r_{t t}=\frac{M}{r^2 c} \big(1-\frac{2M}{rc^2} \big)$, $\Gamma^\phi_{\phi r}=\frac{1}{r}$, and $\Gamma^t_{t r}=\frac{M}{r^2 c} \big(1-\frac{2M}{rc^2} \big)^{-1}$.  For the timeless circular curve, the same reasoning holds after taking $\mu=0$.  

\paragraph{}
On the radial geodesic, Eq.\ref{eq48} together with the corresponding geodesic initial condition $X^r=c \big(1-\frac{2M}{r_0c^2} \big)^\frac{1}{2}$ and $X^t=\big(1-\frac{2M}{r_0c^2} \big)^{-\frac{1}{2}}$ yield

\begin{align}
X^r (r)&=  c \big(1-\frac{2M}{r_0c^2} \big)^\frac{1}{2}  \tag{A10}\label{eqA10}\\
X^t  (r)&= \big(1-\frac{2M}{r_0c^2} \big)^\frac{1}{2} \big(1-\frac{2M}{rc^2} \big)^{-1}, \tag{A11}\label{eqA11}
\end{align}

which of course satisfy the geodesic condition $\frac{X^t (r)}{X^r(r)}=\frac{dt}{dr}= \frac{1}{c} \big(1-\frac{2M}{rc^2} \big)^{-1}$ everywhere.  Once again, these are consistent with the transport equations

\begin{align}
\frac{X^r (r)}{dr} + \Gamma^r_{r r} X^r  (r)+  \frac{1}{c} \big(1-\frac{2M}{rc^2} \big)^{-1} \Gamma^r_{t t} X^t  (r) &=   0   \tag{A12}\label{eqA12}\\
\frac{X^t (r)}{dr} + \frac{1}{c} \big(1-\frac{2M}{rc^2} \big)^{-1} \Gamma^t_{t r} X^r  (r)+   \Gamma^t_{r t} X^t  (r) &=   0. \tag{A13}\label{eqA13}
\end{align}

where $\Gamma^r_{r r}=\frac{M}{r^2 c} \big(1-\frac{2M}{rc^2} \big)^{-1}$ and $\Gamma^t_{rt}=\Gamma^t_{tr}=\frac{M}{r^2 c} \big(1-\frac{2M}{rc^2} \big)$.

\paragraph{}
Lastly, the solution 

\begin{align}
X^r (r)&=  X^r \big(1-\frac{2M}{rc^2} \big)^\frac{1}{2} \sinh (\Omega t) - X^t \big(1-\frac{2M}{rc^2} \big)^\frac{1}{2} \cosh (\Omega t)  \tag{A14}\label{eqA14}\\
X^t (r)&=  -X^r \big(1-\frac{2M}{rc^2} \big)^{-\frac{1}{2}} \cosh (\Omega t) + X^t \big(1-\frac{2M}{rc^2} \big)^{-\frac{1}{2}} \sinh (\Omega t)  \tag{A15}\label{eqA15}
\end{align}

extracted from Eq.\ref{eq54} satisfy  the transport equations

\begin{align}
\frac{X^r (r)}{dr} +  \Gamma^r_{t t} X^t  (r) &=   0  \tag{A16} \label{eqA16}\\
\frac{X^t (r)}{dr} +  \Gamma^t_{t r} X^r  (r) &=   0.\tag{A17}\label{eqA17}
\end{align}

with the same values of $\Gamma^r_{t t}$ and  $\Gamma^t_{tr}$

\end{document}